\documentclass[a4paper,11pt]{article}
\usepackage{pos}
\usepackage{graphicx}
\usepackage{gensymb}
\usepackage{hyperref}
\usepackage{subcaption}

\title{A maximum-likelihood-based technique for detecting extended gamma-ray sources with VERITAS}
 \ShortTitle{A maximum-likelihood-based technique in VERITAS}

\author[a,b]{Alisha Chromey}

\affiliation[a]{VERITAS Collaboration}

\affiliation[b]{Iowa State University, Physics and Astronomy Department,\\
Street number, Ames, Iowa, United States of America}

\forColl{VERITAS} 

\emailAdd{achrmy@iastate.edu}

\abstract{Gamma-ray observations ranging from hundreds of MeV to tens of TeV are a valuable tool for studying particle acceleration and diffusion within our galaxy. Supernova remnants, pulsar wind nebulae, and star-forming regions are the main particle accelerators in our local Galaxy. Constructing a coherent physical picture of these astrophysical objects requires the ability to distinguish extended regions of gamma-ray emission, the ability to analyze small-scale spatial variation within these regions, and methods to synthesize data from multiple observatories across multiple wavebands. Imaging Atmospheric Cherenkov Telescopes (IACTs) provide fine angular resolution (<0.1 degree) for gamma-rays above 100 GeV. Typical data reduction methods rely on source-free regions in the field of view to estimate cosmic-ray background. This presents difficulties for sources with unknown extent or those which encompass a large portion of the IACT field of view (3.5 degrees for VERITAS). Maximum-likelihood-based techniques are well-suited for analysis of fields with multiple overlapping sources, diffuse background components, and combining data from multiple observatories. Such methods also offer an alternative approach to estimating the IACT cosmic-ray background and consequently an enhanced sensitivity to largely extended sources. In this proceeding, we report on the current status and performance of a maximum likelihood technique for the IACT VERITAS. In particular, we focus on how our method’s framework employs a dimension for gamma-hadron separation parameters in order to improve sensitivity on extended sources.}

\FullConference{37$^{\rm{th}}$ International Cosmic Ray Conference (ICRC 2021)\\
		July 12th -- 23rd, 2021\\
		Online -- Berlin, Germany}

\begin{document}

\maketitle


\section{Introduction}

Since the earliest decades of the twentieth century, particle physicists and astronomers have asked the question: What are the origin points of astrophysical cosmic-rays? The cosmic-ray spectrum ranges across multiple decades of energy and past $10^{21}$ eV, therefore, multiple source types must be responsible for the cosmic-ray populations. At the highest energies, there must be an astrophysical process which accelerates charged particles. The most probable sources of the highest energy cosmic-rays produced in the Milky Way are supernova remnants (SNRs), pulsar wind nebulae (PWN), and/or superbubbles. Due to interstellar magnetic fields, cosmic-rays are distributed isotropically from the point of view of local observations, therefore, secondary signals, neutrinos and $\gamma$-rays, are observed to determine the astrophysical origins of very high energy cosmic-rays.

Imaging atmospheric Cherenkov telescopes (IACTs) reconstruct $\gamma$-rays in the energy range of hundreds of GeV to tens of TeV by measuring Cherenkov light produced by extensive air showers. The \textit{Fermi}-Large Area Telescope (LAT) and the High Altitude Water Cherenkov (HAWC) telescope also detect $\gamma$-ray emission, below (peak sensitivity $\sim$1GeV\cite{fermiFGES}) and above (peak sensitivity $\sim$1TeV\cite{hawc2nd}) IACTs' energy range, respectively. Both have detected multiple extended $\gamma$-ray sources associated with SNRs, superbubbles, and PWN. The spectral parameters, significances, and fluxes reported from some of these \textit{Fermi}-LAT and HAWC sources strongly suggest that they are detectable by IACTs.  However, currently operating IACTs have much smaller fields of view, for instance, VERITAS has a 3.5$^{\circ}$ field of view (FOV) diameter. To account for the background events in a standard data reduction, background subtraction is performed between a region centered on a $\gamma$-ray source and an equivalent background region. This approach requires a significant portion of the VERITAS FOV to be free from $\gamma$-ray emission associated with sources of interest. Therefore, straightforward background subtraction is increasingly difficult with increasing size of source extension. The often unknown nature of the extended source morphology is a further complication, as appropriate background regions cannot be defined apriori. A different approach must be implemented for VERITAS data taken on extended sources.

The work presented in this proceeding details a maximum-likelihood method (MLM), implemented through the $\gamma$-ray astronomy analysis package Gammapy\footnote{Gammapy 0.18.2}. For a given set of parameters, both the VERITAS cosmic-ray background and $\gamma$-ray emission are modeled and fit to the data. Due to its wide utility as a Python package and future application with Cherenkov Telescope Array (CTA), Gammapy is the package chosen to accept both reduced data and instrument response functions as inputs and perform the likelihood analysis\cite{gammapy2}\cite{gammapy3}.

\section{VERITAS}

VERITAS (\textbf{V}ery \textbf{E}nergetic \textbf{R}adiation \textbf{I}maging \textbf{T}elescope \textbf{A}rray \textbf{S}ystem) is an array of four 12-meter IACTs located at the Fred Lawrence Whipple Observatory (FLWO) in southern Arizona (31 40N, 110 57W,  1.3km a.s.l.). Each telescope has 345 facets and a camera of 499 photomultiplier tubes at the focal plane. They operate in the energy range from 100 GeV to >30 TeV, with an energy resolution between 15-25\% and an angular resolution <0.1 deg at 1 TeV for 68\% containment. The array can detect flux at the level of 1\% Crab in $\sim$25 hrs with a pointing accuracy error < 50 arc-seconds. For full details of VERITAS and its performance see\cite{nahee}.

\section{3D Maximum Likelihood Method}

The likelihood, L, is the probability that a model of an emission region predicts the parameters of a given data set. The model parameters are optimized by maximizing the likelihood, done by a minimizer that varies the model parameters. Often in a likelihood analysis, the log of the likelihood is calculated since this turns the product into a computationally efficient sum. A general likelihood equation, the product of all probability distribution functions (PDFs) modeling the data, is the following:

\begin{equation}
L=\prod_{i=1}^{N} P_{i}
\end{equation}

where

\begin{equation}
P_{i} = P(\Vec{x};a_1, a_2,...a_m)
\end{equation}

and $\Vec{x}$ represents the function's independent variables. Finding the parameter values that maximize L is the same as determining the parameter values of the modeling equation which are most consistent with the data\cite{MLMref}. 

The 3D-MLM detailed in this proceeding optimizes model parameters to determine the morphology and spectra of extended sources for VERITAS observations taken after September 2009\cite{JCardenzana}. For a set of conditions, instrument response functions (IRFs) model the instrument performance and model the expected number of detected events, given a sky flux and integration time. The IRF function for modeling the spatial distribution of $\gamma$-rays is

\begin{equation}
\label{eq:irfs}
R(p,E|p_{true},E_{true})=A_{eff}(p_{true},E_{true}) PSF(p|p_{true},E_{true}) E_{disp}(E|p_{true},E_{true})
\end{equation}

where:

\begin{description}
\item[$\bullet$]\textbf{$A_{eff}(p_{true},E_{true})$:} Effective Area
\item[$\bullet$]\textbf{$PSF(p|p_{true},E_{true})$:} Point Spread Function
\item[$\bullet$]\textbf{$E_{disp}(E|p_{true},E_{true})$:} Energy Dispersion
\end{description}

In this formalization, $p$ is the reconstructed direction, $p_{true}$ is the true direction, $E$ is the reconstructed energy, and $E_{true}$ is the true energy\footnote{I used the formulation implemented in Gammapy 0.18.2 and detailed at \url{https://docs.gammapy.org/0.18.2/irf/index.html}}. Additionally, though it is not noted in \hyperref[eq:irfs]{Equation 3}, there is also correlation between the IRFs and $\gamma$-hadron separation parameters; therefore a shower parameter, mean scaled width (MSW), is also a dependent parameter is this formulation\cite{citeicrc}. To account for this dependency, a binned-likelihood is performed with events separated into classes by MSW, in addition to binning in reconstructed energy and two bins of camera offset. The separation among MSW is currently 0.8 < MSW < 1.1 and 1.1 < MSW < 1.3. We may use finer binning in the future.

By design, this 3D-MLM analysis fits a spectrum from 316 GeV to 5.01 TeV. Events below 316 GeV are removed to avoid the increased energy bias near the event threshold. To work with a narrower point spread function and achieve better energy resolution, simultaneous detections of atmospheric showers made with only two of the four telescopes in the array are removed\cite{JCardenzana}. 

\subsection{Background Model}
Even after data reduction, emission is dominated by hadronic induced air showers. In order to model the cosmic-ray background, the 3D-MLM uses two parameters sensitive to differences in the shower shape between hadrons and $\gamma$-rays: mean scaled width (MSW) and mean scaled length (MSL). These two parameters are already used in IACT data analysis to separate background emission (cosmic-rays) from $\gamma$-ray emission. For a given pointing (zenith and azimuth), the MSW and MSL of an air shower are derived respectively from the width and length of the shower image:

\begin{equation}
MSX = \frac{1}{N_{tel}} \sum_{i}^{N_{tel}} \frac{x_i}{\langle x_{sim}(size_i , D_i)\rangle} 
\end{equation}

where $N_{tel}$ is the total number of telescopes that have images, \textit{i} is the telescope number, and $\langle ~x_{sim}(size_i , D_i)\rangle$ is the mean width or length of the shower image in a lookup-table of simulations for a given image \textit{size} and impact distance D.

\begin{figure*}[h!]
\centering
\begin{subfigure}{0.48\textwidth}
  \centering
  \includegraphics[width=.95\linewidth]{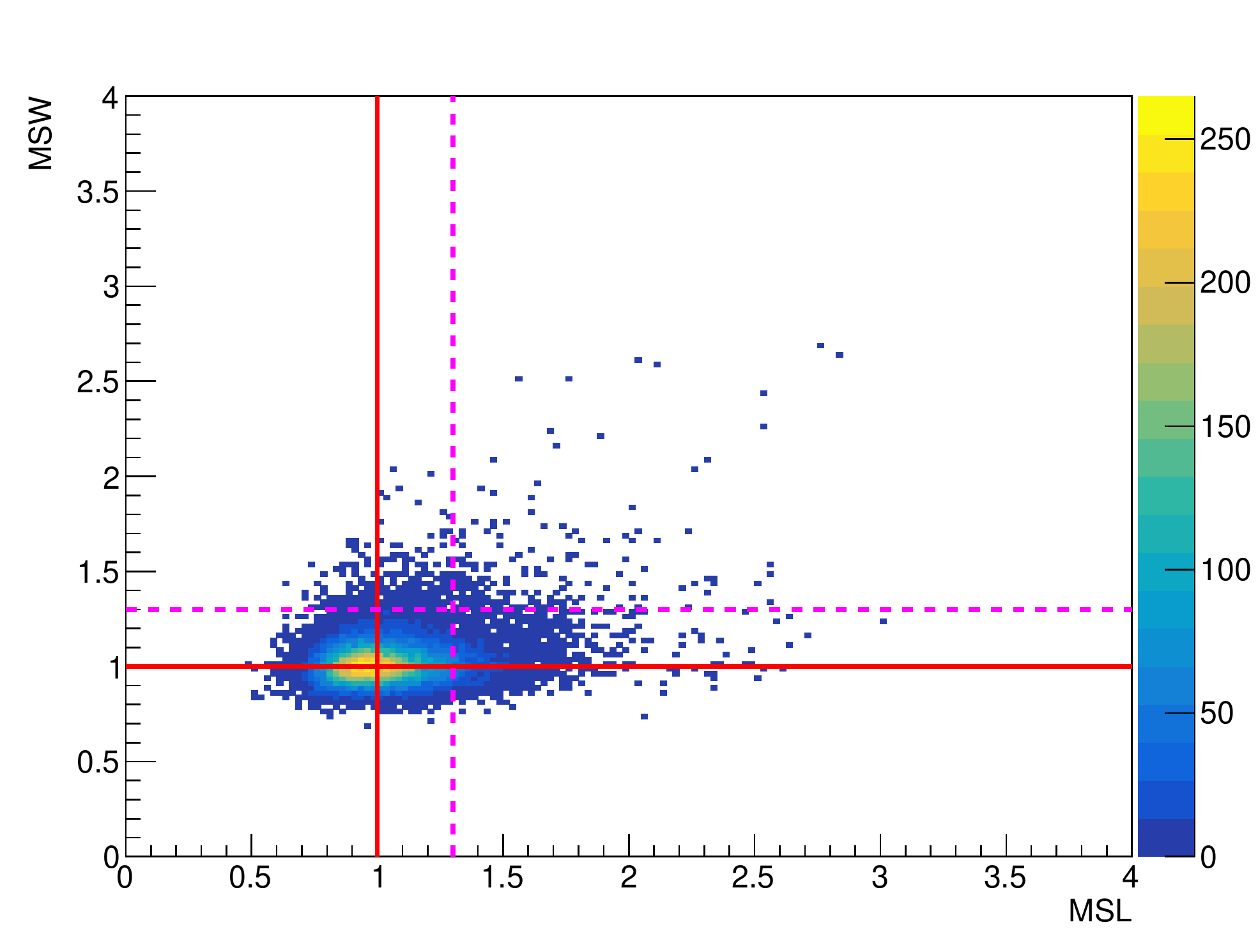}
  \caption{Gamma-ray MSL vs MSW Distribution}
  \label{fig:gammamslmsw}
\end{subfigure}
\begin{subfigure}{0.48\textwidth}
  \centering
  \includegraphics[width=.95\linewidth]{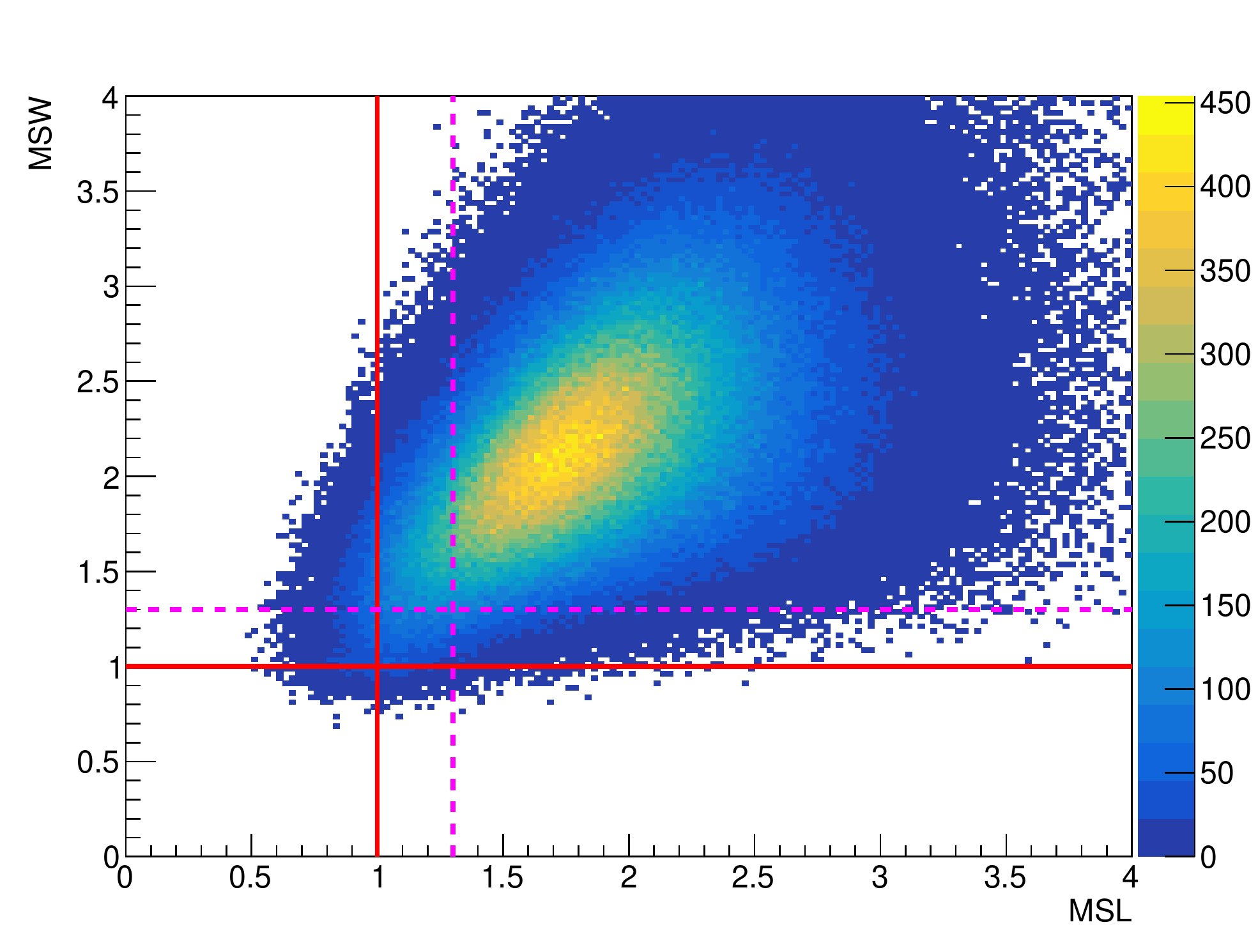}
  \caption{Background MSL vs MSW Distribution}
  \label{fig:segue1mslmsw}
\end{subfigure}
\caption{MSL versus MSW distributions for (a)$\gamma$-ray simulations and (b) events from Segue 1, a background field and dark matter target. The simulations are for observations at $70^{\circ}$ elevation, $180^{\circ}$ azimuth, and $0.5^{\circ}$ offset. The background events are extracted for observations between $75^{\circ}$  to $65^{\circ}$ elevation. Only events with three or four telescopes participating in the shower reconstruction are considered.}
\label{fig:mslmsw}
\end{figure*}

For $\gamma$-ray events, MSW and MSL values peak close to $1$. Hadron events, on the other hand, largely tend to have MSW and MSL values greater than 1.3. Figure \ref{fig:mslmsw} shows that the distributions of MSW and MSL are very different for $\gamma$-ray events and hadron events. A standard analysis for VERITAS data selects MSW values from 0.05-1.1 and MSL from 0.05-1.3. The 3D-MLM incorporates MSW values up to 1.3 to better constrain the shape of the background distribution.

Simulations are used to derive the MSL and MSW distributions of $\gamma$-rays and IRFs. The background emission model is derived from observations of $\gamma$-ray-quiet FOVs or low flux point sources. To account for any potential $\gamma$-ray sources in the selected background samples, bins within 0.4$^\circ$ of source positions are excluded. Similar exclusion regions are applied to the positions of bright stars in the FOV. 

The shapes of the MSL and MSW distributions and IRFs also depend on zenith, azimuth, camera offset, energy, and camera noise. To control for these dependencies, background data samples are divided between bins of these parameters and then the maximum likelihood fit is performed in each set of bins. 

The camera noise needs to be handled with additional care. Often times, the average camera noise in a source field is greater than the average noise in a background field. This is largely because extended sources are usually found in the galactic plane, bright with stars, and most background samples are observed off the galactic plane. There is an algorithm in the VERITAS standard software that introduces artificial noise into data samples, which is also used in the 3D-MLM to match noise levels between source and background fields\cite{fegan}.

A likelihood analysis binned in multiple dimensions reduces the number of available statistics for each background model from a limited sample size. One solution is to convert the 2D MSL vs MSW histogram into a matrix and apply a dimensionality reduction method. The plan going forward is to use singular value decomposition to diagonalize the covariance matrix of MSL vs MSW and study the dependence of the eigenvalues and eigenvectors on energy and observation parameters.

\subsection{Point Spread Function}

Following from previous work, the VERITAS $\gamma$-ray point spread function (PSF) is modeled in the 3D-MLM with the King function \cite{king}\cite{kingfermi}.

\begin{equation}
PSF(x,y) \propto \Big(1- \frac{1}{\lambda} \Big) \Big[ 1+ \Big(\frac{1}{2\lambda}\Big) \cdot \Big(\frac{x^2}{\sigma_{x}^2}+\frac{y^2}{\sigma_{y}^2}\Big) \Big] ^{-\lambda}
\end{equation}.

Previously, a single fit of the spatial distribution was performed for all MSW values. However, in bright source sky maps, such as the Crab, the sky maps show a residual bias at the location of the point source after model subtraction. For the MSW values between 1.1 and 1.3, the modeled PSF overestimates the core emission and underestimates the tail emission. The pattern is reversed for MSW values < 1.1. To address the PSF dependence on MSW, the IRFs are currently binned in two ranges of MSW, a background dominated regime (1.1-1.3), and a $\gamma$-ray dominated regime (0.8-1.1)\cite{citeicrc}.

The initial validation studies were performed assuming a symmetric King function, where $\sigma_{x} = \sigma_{y}$. In previous studies, allowing the $\lambda$ parameter to vary led to instability in extrapolated values, therefore it has a fixed value for all analyses in this proceeding\cite{JCardenzana}.

\begin{figure*}
\centering
\begin{subfigure}{0.48\textwidth}
  \centering
  \includegraphics[width=.95\linewidth]{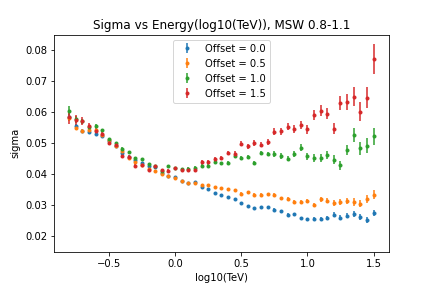}
  \caption{Gamma-ray dominated regime}
  \label{fig:MSW0p8}
\end{subfigure}
\begin{subfigure}{0.48\textwidth}
  \centering
  \includegraphics[width=.95\linewidth]{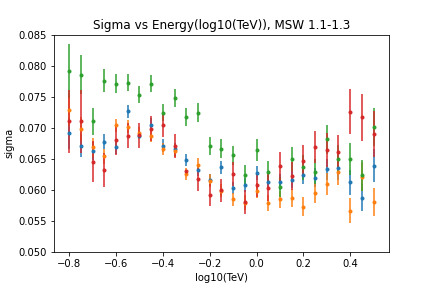}
  \caption{Background dominated regime}
  \label{fig:MSW1p3}
\end{subfigure}
\caption{Results of $\sigma$ versus energy after fit of symmetric King function to $\gamma$-ray simulations generated at zenith=$20^{\circ}$, all azimuths, and pedestal variance=6.75.}
\label{fig:sigmavsE}
\end{figure*}

A series of trials fitting the symmetric King function to the 2D spatial distribution of $\gamma$-ray simulations were performed. The plots in Figure \ref{fig:sigmavsE} show the distribution of $\sigma$ versus $log_{10}(E_{true})$, for four different wobble offsets. In the background dominated MSW regime, shown in Figure \ref{fig:MSW1p3}, the PSF shows no dependence on offset. This is likely due to the overall poorer spatial reconstruction at these MSW values, which drowns out all other dependencies. In the $\gamma$-ray dominated regime of MSW, shown in Figure \ref{fig:MSW0p8}, the PSF shape remains consistent across offset below 1 TeV. Above 1 TeV and for offsets smaller than 0.75$^{\circ}$, $\sigma$ continues to decrease with increasing energy. For offsets above 0.75$^{\circ}$, $\sigma$ increases with larger energy.

\begin{figure*}
\centering
\begin{subfigure}{0.48\textwidth}
  \centering
  \includegraphics[width=.95\linewidth]{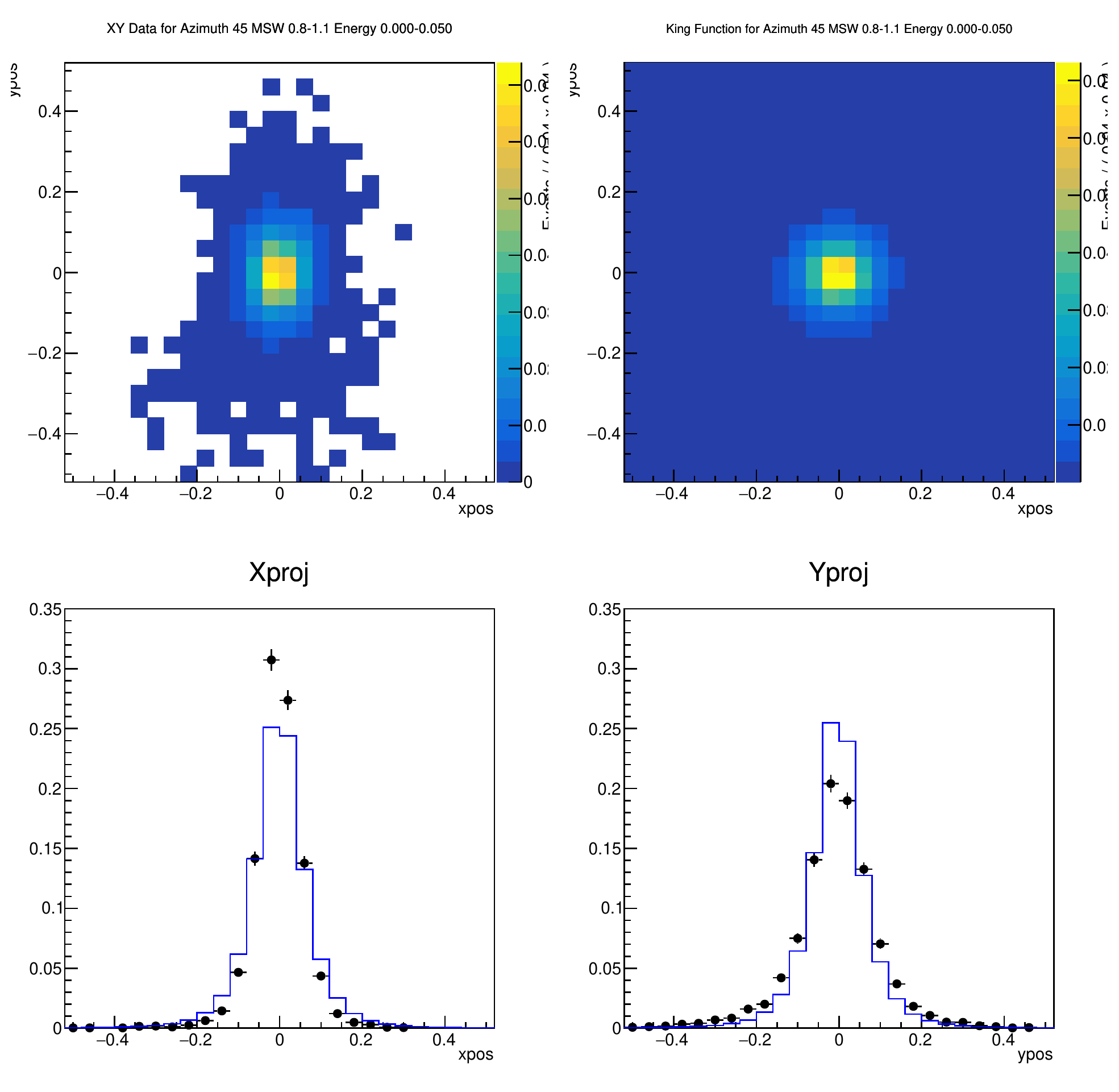}
  \caption{Fit with symmetric King function}
  \label{fig:symmetricfit}
\end{subfigure}
\begin{subfigure}{0.48\textwidth}
  \centering
  \includegraphics[width=.95\linewidth]{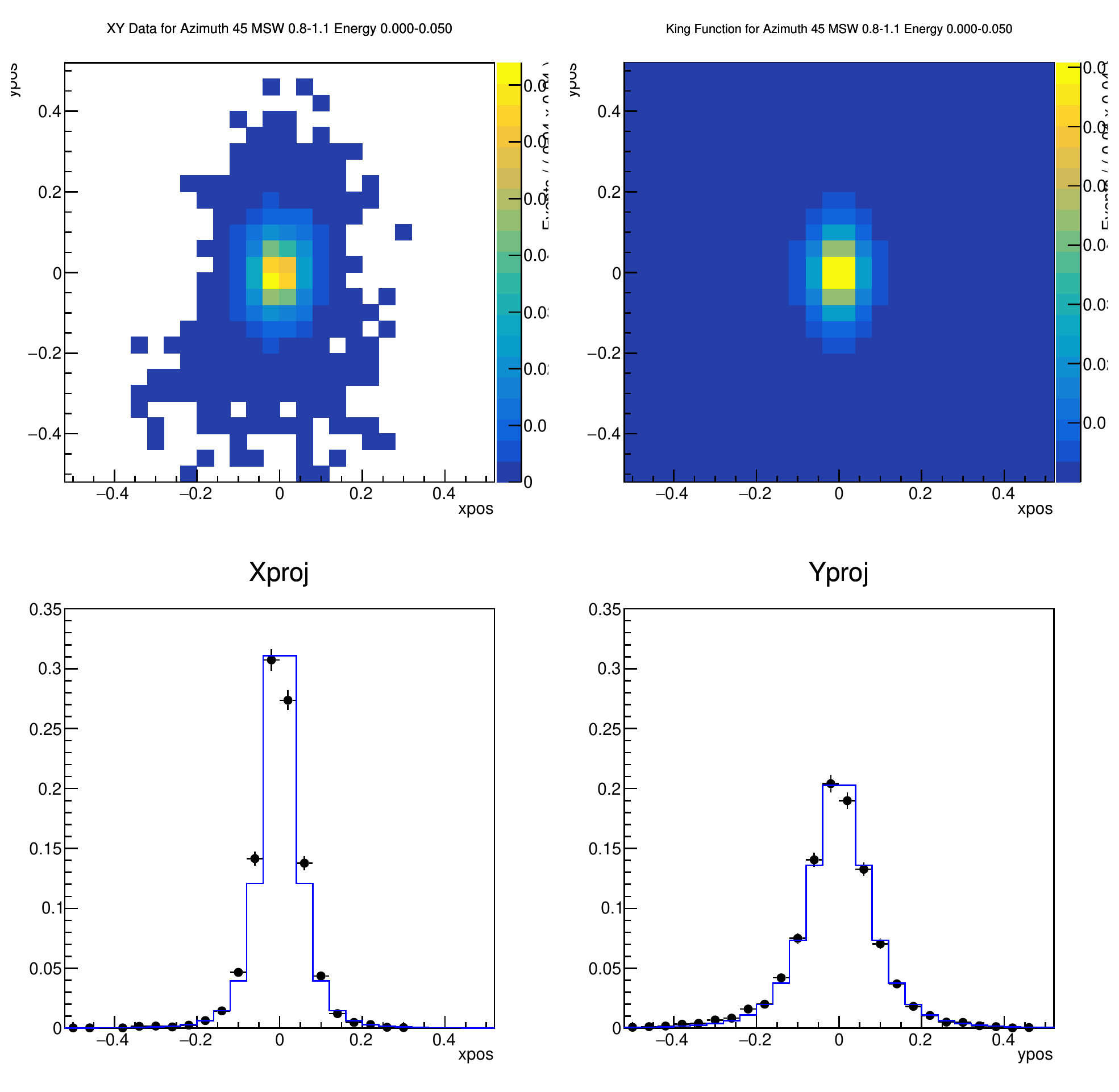}
  \caption{Fit with asymmetric King function}
  \label{fig:asymmetricfit}
\end{subfigure}
\caption{Plots of the King function after fitting and the 2D spatial distribution of simulated events at 1 TeV, zenith=$20^{\circ}$, all azimuths, 1.5$^{\circ}$ offset, and pedestal variance=6.75. In the bottom 1D projection plots, black dots are data and the overlay in blue is a histogram generated from the fitted King function.}
\label{fig:symVSasym}
\end{figure*}

\begin{figure*}
\centering
\begin{subfigure}{0.48\textwidth}
  \centering
  \includegraphics[width=.95\linewidth]{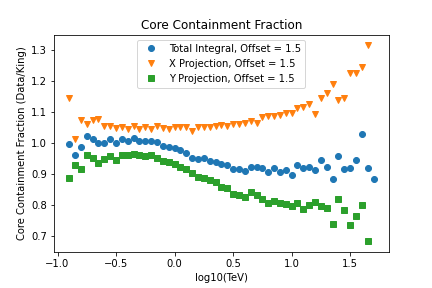}
  \caption{Fit with symmetric King function}
  \label{fig:intsymmetricfit}
\end{subfigure}
\begin{subfigure}{0.48\textwidth}
  \centering
  \includegraphics[width=.95\linewidth]{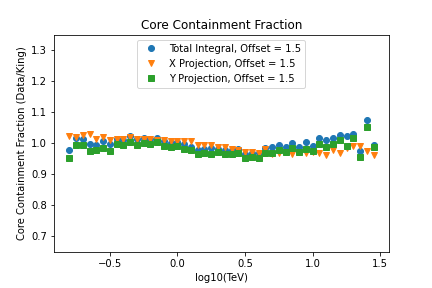}
  \caption{Fit with asymmetric King function}
  \label{fig:intasymmetricfit}
\end{subfigure}
\caption{These plots present the integrated core region of the PSF, compared between the distribution of 1.5$^{\circ}$ offset simulations and the King function shape, after fitting. The integrated difference is plotted versus energy, for the 2D integrated difference and each integrated projection.}
\label{fig:int_symVSasym}
\end{figure*}

Comparing the shape of both the PSF and King function after the fit, for example in Figure \ref{fig:symVSasym}, shows that the broadening of the PSF at larger offsets is not symmetric. The PSF increases along the direction of offset relative to the tracking position. For the set of simulations used to derive IRFs, the tracking position is always in one direction. Changing to an asymmetric King function and fitting for two free parameters, $\sigma_{x}$ and $\sigma_{y}$, improves the quality of the fit at larger offsets. Figure \ref{fig:int_symVSasym} shows the integrated difference between the PSF of data versus the fitted shape of the King function, for both the symmetric and asymmetric fits. When the King function is in asymmetric form, the integrated difference is close to 1 and nearly similar for both the entire 2D shape and each projection, further demonstrating that the asymmetric King function is the better model for the PSF in all cases. Therefore, an asymmetric King function with a fixed value for $\lambda$ works best for a spatial distribution model applied to VERITAS data.

\section{Validation Sources}
Once all the IRFs and background modeling have been tested a series of validation tests will be performed with the full 3D-MLM. Validation will be done on bright point-like emission and standard sources, such as the Crab. Most importantly, a search for null results from a 3D-MLM analysis will be performed on dark matter targets, such as, Segue 1 and Ursa Minor. Finally, the new analysis will be tested on known extended sources that have already been studied and published with standard analysis, for example, slightly extended PWN IC 443.

\section{Conclusions}
Multiple extended galactic $\gamma$-ray sources have yet to be observed with IACTs such as VERITAS, even in cases where the projected flux levels and spectral indices reach VERITAS sensitivities. This is due to the difficulty subtracting background emission from a FOV of only a few degrees across and filled with source emission of unknown morphology. A maximum-likelihood method for VERITAS observations in under development. In addition to the spatial model, the background and source models utilize the shower parameters, MSL and MSW, as dimensions for discriminating background and source emission. Results of past validation studies of the 3D-MLM on the brightest $\gamma$-ray point sources show a dependence of PSF on MSW that is in the process of being resolved. Going forward, this is taken into account by characterizing the PSF in two ranges of MSW. The asymmetric King function has also been shown as the best function to model the $\gamma$-ray PSF. Validations studies of background models for the 3D-MLM are ongoing.

Successful observations of extended sources in the very high energy range will contribute to determination of the composition of accelerated particle populations and $\gamma$-ray production mechanisms of supernova remnants, such as IC 443 and Gamma Cygni, and TeV halos, such as Geminga. 

\acknowledgments
This research is supported by grants from the U.S. Department of Energy Office of Science, the U.S. National Science Foundation and the Smithsonian Institution, and by NSERC in Canada. This research used resources provided by the Open Science Grid, supported by the National Science Foundation and the U.S. Department of Energy's Office of Science, and resources of the National Energy Research Scientific Computing Center (NERSC), a U.S. Department of Energy Office of Science User Facility operated under Contract No. DE-AC02-05CH11231. We acknowledge the excellent work of the technical support staff at the Fred Lawrence Whipple Observatory and at the collaborating institutions in the construction and operation of the instrument. Dr. Amanda Weinstein and Alisha Chromey acknowledge the support from grant NSF-PHY 1555161.

\clearpage \section*{Full Authors List: \Coll\ Collaboration}

\scriptsize
\noindent
C.~B.~Adams$^{1}$,
A.~Archer$^{2}$,
W.~Benbow$^{3}$,
A.~Brill$^{1}$,
J.~H.~Buckley$^{4}$,
M.~Capasso$^{5}$,
J.~L.~Christiansen$^{6}$,
A.~J.~Chromey$^{7}$, 
M.~Errando$^{4}$,
A.~Falcone$^{8}$,
K.~A.~Farrell$^{9}$,
Q.~Feng$^{5}$,
G.~M.~Foote$^{10}$,
L.~Fortson$^{11}$,
A.~Furniss$^{12}$,
A.~Gent$^{13}$,
G.~H.~Gillanders$^{14}$,
C.~Giuri$^{15}$,
O.~Gueta$^{15}$,
D.~Hanna$^{16}$,
O.~Hervet$^{17}$,
J.~Holder$^{10}$,
B.~Hona$^{18}$,
T.~B.~Humensky$^{1}$,
W.~Jin$^{19}$,
P.~Kaaret$^{20}$,
M.~Kertzman$^{2}$,
T.~K.~Kleiner$^{15}$,
S.~Kumar$^{16}$,
M.~J.~Lang$^{14}$,
M.~Lundy$^{16}$,
G.~Maier$^{15}$,
C.~E~McGrath$^{9}$,
P.~Moriarty$^{14}$,
R.~Mukherjee$^{5}$,
D.~Nieto$^{21}$,
M.~Nievas-Rosillo$^{15}$,
S.~O'Brien$^{16}$,
R.~A.~Ong$^{22}$,
A.~N.~Otte$^{13}$,
S.~R. Patel$^{15}$,
S.~Patel$^{20}$,
K.~Pfrang$^{15}$,
M.~Pohl$^{23,15}$,
R.~R.~Prado$^{15}$,
E.~Pueschel$^{15}$,
J.~Quinn$^{9}$,
K.~Ragan$^{16}$,
P.~T.~Reynolds$^{24}$,
D.~Ribeiro$^{1}$,
E.~Roache$^{3}$,
J.~L.~Ryan$^{22}$,
I.~Sadeh$^{15}$,
M.~Santander$^{19}$,
G.~H.~Sembroski$^{25}$,
R.~Shang$^{22}$,
D.~Tak$^{15}$,
V.~V.~Vassiliev$^{22}$,
A.~Weinstein$^{7}$,
D.~A.~Williams$^{17}$,
and 
T.~J.~Williamson$^{10}$\\
\noindent
$^1${Physics Department, Columbia University, New York, NY 10027, USA}
$^{2}${Department of Physics and Astronomy, DePauw University, Greencastle, IN 46135-0037, USA}
$^3${Center for Astrophysics $|$ Harvard \& Smithsonian, Cambridge, MA 02138, USA}
$^4${Department of Physics, Washington University, St. Louis, MO 63130, USA}
$^5${Department of Physics and Astronomy, Barnard College, Columbia University, NY 10027, USA}
$^6${Physics Department, California Polytechnic State University, San Luis Obispo, CA 94307, USA} 
$^7${Department of Physics and Astronomy, Iowa State University, Ames, IA 50011, USA}
$^8${Department of Astronomy and Astrophysics, 525 Davey Lab, Pennsylvania State University, University Park, PA 16802, USA}
$^9${School of Physics, University College Dublin, Belfield, Dublin 4, Ireland}
$^{10}${Department of Physics and Astronomy and the Bartol Research Institute, University of Delaware, Newark, DE 19716, USA}
$^{11}${School of Physics and Astronomy, University of Minnesota, Minneapolis, MN 55455, USA}
$^{12}${Department of Physics, California State University - East Bay, Hayward, CA 94542, USA}
$^{13}${School of Physics and Center for Relativistic Astrophysics, Georgia Institute of Technology, 837 State Street NW, Atlanta, GA 30332-0430}
$^{14}${School of Physics, National University of Ireland Galway, University Road, Galway, Ireland}
$^{15}${DESY, Platanenallee 6, 15738 Zeuthen, Germany}
$^{16}${Physics Department, McGill University, Montreal, QC H3A 2T8, Canada}
$^{17}${Santa Cruz Institute for Particle Physics and Department of Physics, University of California, Santa Cruz, CA 95064, USA}
$^{18}${Department of Physics and Astronomy, University of Utah, Salt Lake City, UT 84112, USA}
$^{19}${Department of Physics and Astronomy, University of Alabama, Tuscaloosa, AL 35487, USA}
$^{20}${Department of Physics and Astronomy, University of Iowa, Van Allen Hall, Iowa City, IA 52242, USA}
$^{21}${Institute of Particle and Cosmos Physics, Universidad Complutense de Madrid, 28040 Madrid, Spain}
$^{22}${Department of Physics and Astronomy, University of California, Los Angeles, CA 90095, USA}
$^{23}${Institute of Physics and Astronomy, University of Potsdam, 14476 Potsdam-Golm, Germany}
$^{24}${Department of Physical Sciences, Munster Technological University, Bishopstown, Cork, T12 P928, Ireland}
$^{25}${Department of Physics and Astronomy, Purdue University, West Lafayette, IN 47907, USA}

%
%
%

\end{document}